\documentclass[12pt,preprint]{aastex}

\newcommand{\hst}{{\sl HST}}
\newcommand{\omcen}{{$\omega$ Cen}}

\slugcomment{ApJL Accepted, March 4$^{\rm th}$, 2004}

\shorttitle{The \omcen\ puzzle}
\shortauthors{Bedin et al.}

\begin{document}

\title{OMEGA     CENTAURI:      THE     POPULATION     PUZZLE     GOES
DEEPER\footnote{Based  on observations with  the NASA/ESA  {\it Hubble
Space Telescope},  obtained at the Space  Telescope Science Institute,
which is operated by AURA, Inc., under NASA contract NAS 5-26555.}}

\author{Luigi R. Bedin\altaffilmark{2},        
        Giampaolo Piotto\altaffilmark{2},     
	Jay Anderson\altaffilmark{3},    
	Santi Cassisi\altaffilmark{4},    
	Ivan  R. King\altaffilmark{5},   
	Yazan Momany\altaffilmark{2}, and 
	Giovanni Carraro\altaffilmark{2}}

\altaffiltext{2}{Dipartimento  di Astronomia, Universit\`a  di Padova,
Vicolo     dell'Osservatorio     2,     I-35122     Padova,     Italy;
bedin-piotto-momany-carraro@pd.astro.it}

\altaffiltext{3}{Department of  Physics and Astronomy,  Mail Stop 108,
Rice    University,   6100   Main    Street,   Houston,    TX   77005;
jay@eeyore.rice.edu}

\altaffiltext{4}{Osservatorio Astronomico di Collurania, via M. Maggini,
64100 Teramo; cassisi@astrte.te.astro.it}

\altaffiltext{5}{Department of Astronomy, University of Washington,
Box 351580, Seattle, WA 98195-1580; king@astro.washington.edu}

\begin{abstract}
We present \hst\ observations that show a bifurcation of colors in the
middle main  sequence of the globular cluster   $\omega$ Centauri.  We
see  this in three different  fields,  observed with different cameras
and filters.  We also  present high precision  photometry of a central
ACS field which shows a  number of main-sequence turnoffs and subgiant
branches.  The double  main  sequence,   the  multiple turnoffs    and
subgiant  branches, and other population   sequences discovered in the
past along    the red  giant branch   of   this cluster  add   up to a
fascinating but frustrating puzzle.   We suggest various explanations,
none of them very conclusive.
\end{abstract}


%
\section{Introduction}
%

A number of  properties (total mass, chemical composition, kinematics,
and spatial   distribution of  the  stars)  make $\omega$  Centauri  a
peculiar object among Galactic   globular clusters.  The  most evident
anomaly is the large spread in metallicity  seen both in spectroscopic
(Norris \&  Da Costa 1995)  and photometric (Hilker \&  Richtler 2000,
Lee et al.\ 1999, Pancino et al.\ 2000) investigations.

Most of  the  fascinating results   on \omcen~  come  from the evolved
stellar  population, which can be studied   in detail from the ground.
In this paper we use \hst~ data to explore  the cluster's turnoff (TO)
and main-sequence (MS) populations.  While studies  of the evolved red
giant  branch    (RGB)  can   explore   metallicity,   kinematic,  and
spatial-distribution issues, we  need the fainter stars  if we hope to
learn anything about ages  and mass functions, and  to give us  better
statistics (there are $\sim$10 MS stars for every RGB star).

The present  paper was stimulated by preliminary  results by one of us
(Anderson 1997, 2002, 2003), on the presence  of multiple turnoffs and
of a bifurcated main sequence (MS).  Here we  confirm that the unusual
features    found  in   the  color-magnitude  diagrams  are  not  some
data-reduction artifact, or a local phenomenon.  The features are real
and  are present throughout the  cluster.  Unfortunately, the striking
results  we present  here lead  to more  questions   than they answer.
Though it will take  a lot of time  to fully exploit the  \omcen~ data
stored in the \hst~ archive  (and we  are  working on this), we  think
that these new results are  worthy of immediate publication because of
their importance to the ongoing debate on the nature of this object.

%
\section{Observations and Data Reduction}
%

In  this paper   we  use  WFPC2  and   ACS  \hst~  data to   construct
color--magnitude   diagrams (CMDs) which   extend from 1--2 magnitudes
above the TO  to more than 7  magnitudes below it.  In particular,  we
have used the following sets of images:
1) GO 9444 (ACS/WFC):\  4$\times$1350s  F814W and  4$\times$1350s  F606W;
2) GO 9442 (ACS/WFC):\ 27$\times$340s   F435W and 36$\times$440s   F658N
(a $3\times3$ mosaic at the center);
3) GO 6821 (WFPC2):\   2$\times$1s, 2$\times$10s, 8$\times$100s in F675W,
   and   26s,  2$\times$260s in   F336W;
4) GO 5370 (WFPC2):\ 2$\times$300s, 600s F606W, and 2$\times$400s, 1000s
   in F814W.
All these  images have been reduced with  the  algorithms described in
Anderson \& King  (2000).     Photometric calibration has been    done
according to the  Holtzman et al.\  (1995) flight system for the WFPC2
data; for   ACS/WFC we used the   preliminary Vega System  zero points
available on the ACS website.

%
\section{The color--magnitude diagram: observational evidence}
%

The left part of Figure 1 shows four CMDs:  the two upper panels focus
on the turnoff region, and the lower ones on the main-sequence region.
Panel a  shows the original WFPC2 CMD  that first discovered the lower
turnoff   (LTO)  sequence (Anderson 2003).    Panel  b shows  the same
sequence (from WFC data) with many more details and more stars, from a
larger region of the cluster.  There are a number  of distinct TOs and
subgiant  branches (SGBs), and   the  connection of   the LTO  to  the
metal-rich red giant branch (called RGB-a by Pancino et al.\ 2000) can
be  seen more clearly with  more stars.   We show H$\alpha$ photometry
here instead of $R_{625}$ because there  are more exposures and we can
therefore make a better estimate of the photometric errors.

Ferraro  et al.\  (2004) reduced the   same ACS field in their  recent
study of the LTO  population.  They made  the obvious association that
this LTO  corresponds  to the  metal-rich ([Fe/H] $\sim\!-0.5$)  RGB-a
population, and found that  it could be fit only  with an old (15 Gyr)
isochrone.   They   found little  or  no  age   difference between the
metal-rich and metal-poor  populations.  In our  diagram,  it is clear
that not  only is there a  lower subgiant branch,  but we can also see
the  lower turnoff at  a  color that is   clearly redward of the  main
population  sequence.  We  also  see  several SGBs:  the  LTO   SGB, a
distinct SGB  at   the bright end of  the   main population, and  some
distinct structure  in the  region between the   upper and lower  SGB.
Clearly, this is telling us a lot about the cluster's populations, but
the  interpretation is complicated   by age, metallicity, and distance
degeneracies.
We  will confine the focus of  this  letter to the populations farther
down  the main-sequence, deferring a detailed  analysis of the turnoff
population to when all the WFC central data have been reduced and some
information  on the  metallicity of  the  different   TO-SGBs will  be
available.

Surely  the most  intriguing feature in  the CMDs   of Fig.\ 1  is the
double main sequence (DMS), which is clearly visible in the bottom two
panels   (c,       d).      Panel     c    shows      the     original
$V_{606}\,$vs.$\,V_{606}\!-\!I_{814}$  WFPC2 CMD  from Anderson (1997,
2002),  where the DMS  was  first identified.   Not  only is  the main
sequence much broader than photometric errors, it appears to bifurcate
into two distinct  sequences, with a  region  between the two that  is
almost    devoid      of    stars.    Panel      d     shows   a   new
$V_{606}\,$vs.$\,V_{606}\!-\!I_{814}$  CMD from  ACS/WFC images, which
also shows the  anomalous DMS in a  different field, at 17$'$ from the
center.  The DMS can even be  discerned in the  very inner part of the
cluster (panels b and e).

%
%

We have   CMDs  from different  fields, observed  with  two  different
cameras, in different photometric bands.  The anomalous DMS is present
in  all four  (panels b, c,  d,  e).  There  can be  no doubt that the
double sequence is a real and ubiquitous feature in \omcen.

If we were to guess what  the main sequence should  look like from our
knowledge of the stars on the giant branch, we would expect a sequence
about  0.03   mag in width,  with  a   concentration to  a  blue edge,
corresponding  to  the  metal-poor (MP)   population  containing about
65$\%$  of the   stars, a   tail to the     red corresponding to   the
intermediate-metallicity (Mint) population  containing about 30$\%$ of
the stars, and a small even redder component from the metal-rich RGB-a
population with 5$\%$  of  the stars (we  adopt  the population labels
from Pancino et al.\ 2000).  The sequence we observe here could not be
more different from these expectations.

Let us take note of a few simple facts from Fig.\ 1:
  (1) The two sequences  are   clearly separated,  at  least  in   
      the interval  $22<V_{606}<20.5$, with a  region almost devoid of
      stars between them.
  (2) The bluer MS  (bMS) is much less populous  than the red MS  (rMS). 
      The  bMS  contains 25$\%$  to  35$\%$ of the stars.
  (3) The DMS extends down to at least $V_{606}\sim23.5$. 
      Below this, the bMS appears to vanish, though it is difficult to
      say for sure if it 
      peters out, or  if it blends with the rMS
      as the photometric errors increase.  
  (4) Finally we note that panels b and e show clearly that 
      the bMS is a different population from the LTO--RGB-a population.

The  DMS that we observe   represents a real  puzzle  for at least two
reasons.
First, the bifurcation itself is puzzling.  As summarized 
  above,  the    many    detailed  photometric     and   spectroscopic
  investigations of the RGB  indicate  a spread of metallicities,  not
  two distinct populations.  The  only truly distinct  population seen
  is the metal-rich component.
Second, the less populous of our two MSs is the blue one.  This is
  even more difficult  to understand.  Assuming that  all the stars in
  the two MSs are members of \omcen, any canonical stellar models with
  canonical chemical  abundances  tell us  that the bMS  {\it must} be
  more  metal poor than  the rMS.  However, both spectroscopic (Norris
  \& Da Costa 1995) and  photometric (e.g.,  Hilker \& Richtler  2000)
  investigations show that   the  distribution in metallicity  of  the
  \omcen~ stars begins with a peak  at [Fe/H] $\!\sim\!-1.6$, and then
  tails off on the metal-rich side.
%

%
\section{Comparison with theoretical models}
%

The  previous  section  has   confronted  us with  several   seemingly
contradictory  observational facts concerning  the CMD and populations
of \omcen.  In this section we will see what  light stellar models can
shed on the situation.  Because the  ACS photometric system is not yet
adequately calibrated,  we will confine our isochrone-fitting analysis
to the WFPC2 data.  The adopted stellar models are an extension of the
updated evolutionary models  for very-low-mass stars and more  massive
ones presented by   Cassisi et  al.\  (1999, 2000).    All models  and
isochrones have been transformed  into the WFPC2 observational  planes
by using the accurate bolometric  corrections kindly provided to us by
F.\ Allard (see also Allard et al.\ 1997).

In fitting   the    stellar  models  we  have   adopted   a  reddening
$E(B-V)=0.13$  and a distance modulus    $(m-M)_0=13.6$.  We used  the
absorption coefficients  for  the WFPC2 bands listed  in  Table 12b of
Holtzman et al.\ (1995).  For the F606W band we adopted  a mean of the
absorption  coefficients in F555W and  F675W.  In Figure  2a and 2c we
have overplotted 4  sets of isochrones, corresponding to metallicities
from [Fe/H] $\!=\!-2.1$ to $-0.6$, which covers the entire metallicity
range of the stellar population of \omcen~  (Norris \& Da Costa 1995),
and corresponding to an age of 14 Gyr.

Since we are able to  see the actual turnoff  and not just the SGB (as
in Ferraro et al.\ 2004), and since we use a  $U\!-\!R\,$ CMD, we are
more sensitive  to the population's  metallicity.  Panel a of  Fig.\ 2
shows    that  a   more    metal-poor isochrone  would  fit    the LTO
better. Indeed, Origlia et al.\ (2003) have shown that the RGB-a stars
have a metallicity in  the interval $-0.9<$ [Fe/H] $<\!-0.5$.  It does
appear that the  LTO population would  be fit with a metallicity  near
the middle of this range.

Panel b of Fig.\ 2 shows the  cluster main sequence  in the F606W vs.\
F606W$-$F814W   bands,  along  with   isochrones   for  the  cluster's
metallicity  range.  There are  several features worth noting:\ first,
the [Fe/H]\   $=\!-1.6$  isochrone clearly  follows   along  the  rMS.
Second, canonical models are not able to explain the bMS.  In order to
explain the 0.06 mag $V-I$ color  difference between the two sequences
at  $V$ $\sim$  21,  we would have to   assume that the  bMS stars are
extremely metal-poor ([Fe/H]   $\ll  \!-$2.0), though   it  would seem
absurd to have  such a large   population of low-metallicity  MS stars
when  there is   no  evidence whatever    for such   stars  along  the
much-studied RGB.  The  only  way to force  the  model to fit  the two
sequences with  the metallicity spread   seen in the  RGB would  be to
shift the models arbitrarily in color and/or magnitude.

%
%

We can imagine four possible explanations, though we admit that all of
them seem far-fetched.
(1) The models or calibrations are grossly in error {\it and} 
    the distribution of metallicities is vastly different for
    the RGB stars than for the MS stars.
(2) The bMS represents some super-metal-poor population 
    ([Fe/H] $\ll-2.0$).
(3) The bMS represents a super-helium-rich ($Y\geq0.3$) population. 
(4) The bMS represents a population of stars about 1--2 kpc behind
\omcen.
We examine these four possibilities in the following section.
  
%
\section{Discussion}
%

One way  to interpret the observations  is  to assume  that either the
photometric  calibration or the isochrones  are  in error.  If the net
error is 0.06  mag in  $V-I$  color, then perhaps  the metal-poor (MP)
population ([Fe/H]\ $=\!-1.6$) follows  along the bMS instead of along
the  rMS.  If this is  the case then the  rMS would  correspond to the
metal-rich  population ([Fe/H]\ $=\!-0.5$), as  the 0.06 mag $V\!-\!I$
separation  cannot be explained  by the metallicity difference between
the  MP and Mint  populations.  (While it may  be conceivable that the
isochrones  could have  errors in an   absolute sense, they  should be
reliable in a differential sense.)  There are additional problems with
this  interpretation.  First is that  only 5$\%$ of  the RGB stars are
metal rich, but in this scenario over 70$\%$  of the MS stars would be
metal  rich.  This would  imply  drastically different mass functions,
such as have never  been seen before  anywhere (see Piotto  \& Zoccali
1999).    Furthermore,  there is  no   actual   gap in  the   observed
metallicity distribution and in the color  distribution of the RGBs of
the MP and  Mint populations.  Most importantly, the  fact that the MS
extension of the LTO runs parallel to the rMS (on the  red side of it,
panel e) makes this scenario impossible.

The second interpretation is that the rMS corresponds to the MP stars,
but the bMS corresponds to  a super-metal-poor population, with [Fe/H]
$\ll\!-2$.  However, such a  large population of metal-poor stars  has
never been observed in \omcen\ or in any other globular.

The third  possibility is  that  the populations of   the two MSs have
sensibly different helium  content ($Y$).  Norris, Freeman, \& Mighell
(1996)  have shown that  the metallicity distribution of \omcen\ stars
can be well  fitted by two separate  components, and argued  that this
can be explained by two successive epochs of star formation.  Assuming
for the more metal-rich  ([Fe/H]\ $=\!-1.0$) Mint population  a helium
content of  $Y\sim0.30$, we  find that the  corresponding  MS would be
$\sim0.07$ magnitude bluer in ($V-I$) than the  MP MS (assumed to have
a canonical $Y=0.23$, and [Fe/H]\ $=\!-1.6$). Note that Norris et al.\
(1996) found that the  ratio of the Mint  to  MP population should  be
0.2, compatible, within the uncertainties, with the  value we find for
the rMS/bMS ratio.  Panel e  of Fig.\ 1 shows that  the bMS could well
be  connected with the intermediate TO-SGB.  Panel  a of Fig.\ 2 shows
that this  intermediate SGB is slightly  brighter  than the luminosity
expected for a metallicity similar to the Mint, and the expected TO is
redder than the observed one. These observational facts are consistent
with this  population beeing helium enhanced  and slightly younger, as
expected  if  the  helium  enhancement  is  due to self-pollution from
intermediate AGB MP   stars.    The dramatic increase    of  s-process
heavy-element   abundances  with metallicity  found   by Smith et al.\
(2000) in \omcen\ RGB stars furtherly support the hypothesis that Mint
stars  could have formed from  material  polluted by ejecta from 1.5-3
$m_\odot$ AGB stars.  The  presence of a  population with  high helium
content could also  account  for the  anomalously  hot HB  of  \omcen,
following  the calculations   of  D'Antona et al.\  (2002).   All this
notwithstanding, a $Y\geq0.30$   is higher   than  any value   so  far
measured   in Galactic GCs  (Salaris  et al.\ 2004),   and not easy to
understand.

As a fourth possibility, if we assume that  the rMS corresponds to the
majority    of the cluster   stars,   the bMS  could  correspond  to a
population of  stars located behind \omcen.   As  shown in panel  d of
Fig.\ 2, if the   bMS is populated by   stars located 1.6  kpc  beyond
\omcen, we can easily fit it with an [Fe/H]  $=\!-1$ isochrone.  Panel
e of Fig.\ 1 appears to strengthen this hypothesis: we see the bMS get
closer and  closer to the  rMS, crossing it at  $H\alpha\sim18.5$, and
apparently continuing into a broadened TO and SGB.  This broadening of
the  intermediate   TO could  be   the  result of  a   spread  in both
metallicity and  distance.  The overall  appearance of the CMD is that
there are two sequences, shifted   by up to $\sim0.3$--0.5  magnitude.
The  hypothesis of a background  agglomerate of stars with metallicity
around  [Fe/H] $\sim\!-1.0$ would also naturally  explain  why the bMS
appears to intersect the rMS at  $V_{606}\sim23.5$ (cf.\ Figs.\ 1c and
1d).  Such a background object would naturally explain the observation
that   the  giants of  different   metallicity appear to have somewhat
different spatial  distributions  (Jurcsik 1998,  Hilker  \&  Richtler
2000), though this  spatial variation could  be explained by merger or
self-enrichment scenarios as well.

Leon, Meylan, \&  Combes  (2000) have  identified a tidal  tail around
\omcen.  Tidal tails often have a  clumpy nature.  However, the number
of stars in the bMS  seems to be too  large and the sequence too sharp
to be interpreted as a  part  of a clump   in a tidal tail behind  the
cluster.  Another possibility is that the object  in the background is
a distinct cluster  or a dwarf  galaxy.  As  it should cover  at least
20--30 arcmin in  the sky (this  is the extent of  the region where we
identified a DMS) and   be located at about 7   kpc from the  Sun, the
object should be extended  by at least  40--60 pc.  The probability of
observing such an object in the direction of \omcen~ is extremely low.
However,  if   this object  happens  to be  gravitationally  linked to
\omcen~(either  because it  was part of  the  same original  system or
because it  is the remnant of some  merging event), that would enhance
the probability of seeing it in the same direction as \omcen.

We note that the idea of a population of  stars behind the cluster has
been suggested before.  Ferraro et al.\  (2002) measured a bulk motion
for  the RGB-a stars   with respect to  the  other cluster  stars, and
interpreted this as evidence that it  could be a background object, or
a  merger product that has  not  yet phase-mixed.  However, Platais et
al.\    (2003) find  this   motion  spurious,  attributing   it   to a
color/magnitude term  in    the proper  motions.  Moreover,   Anderson
(2003), using very accurate WFPC2 proper motions, contradicts the bulk
motions seen by Ferraro et  al.\ (see his  Fig.\ 1).  In any case, the
background population  we consider  here could  not correspond to  the
very  metal-rich population; our Fig.\ 1e  makes it clear that the LTO
and the bMS are not related to each other.

%
\section{Resolving the controversy}
%

Much of the current puzzle stems from our inability to interrelate the
various  RGB, SGB, TO, and  MS populations.   Currently there exists a
good  data  base of    observations for RGB   stars,   but  only  spot
observations of the SGB, TO, and MS populations.  An accurate analysis
of   the  proper-motion, radial-velocity,   metallicity,  and  spatial
distributions of  the MS, TO, SGB  stars  of \omcen, in a  large field
sampled over the inner   20 arcmin or  so  of the cluster,  along with
detailed theoretical calculations, are absolutely essential to explain
the observational  facts,  which at the  moment  represent  a mixed-up
puzzle.  The new results presented in this paper show that the more we
learn about this cluster, the more we realize we do not know.

\acknowledgments

We  thank the referee,  John Norris,  and Raffaele  Gratton for useful
suggestions and discussions on the  role of helium in interpreting the
observations presented in this paper.
L.R.B.,  G.C., S.C., Y.M.,  and G.P.\ acknowledge financial support by
MIUR  (PRIN2001,   PRIN2002,  and   PRIN2003).   J.A.\   and   I.R.K.\
acknowledge support by STScI grant GO 9444.

\clearpage

\begin{figure}
\epsscale{1.00}
\plotone{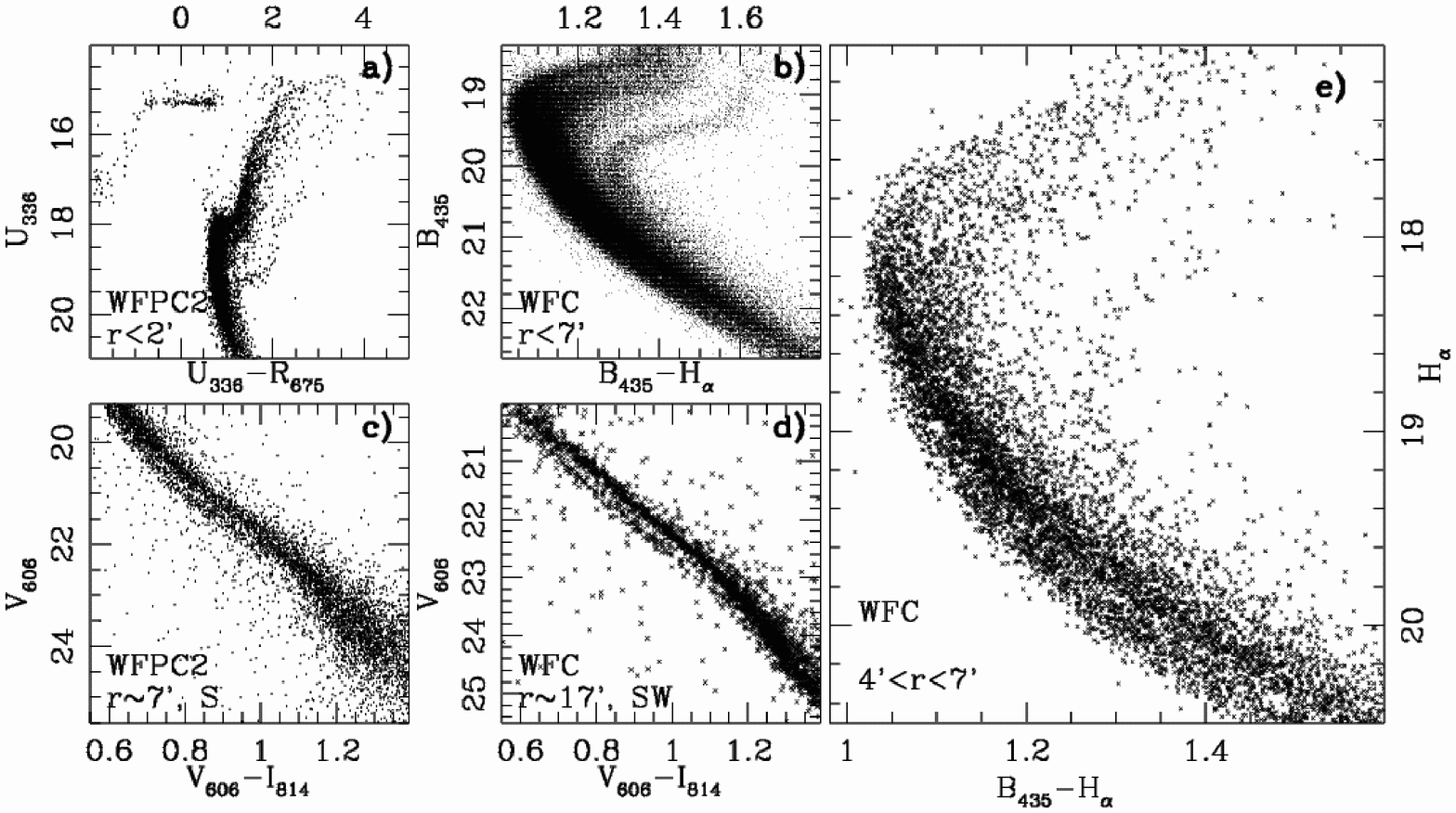}
\caption{A collection of CMDs from  WFPC2 and ACS data of \omcen.  For
each  CMD, the  label  indicates the distances  of the  field from the
cluster  center.  Panel e shows the  subsample of the stars plotted in
panel b,  located at radial  distances r$>4'$ and with photometric rms
lower than 0.025 magnitudes. }
\end{figure}

\clearpage

\begin{figure}
\plotone{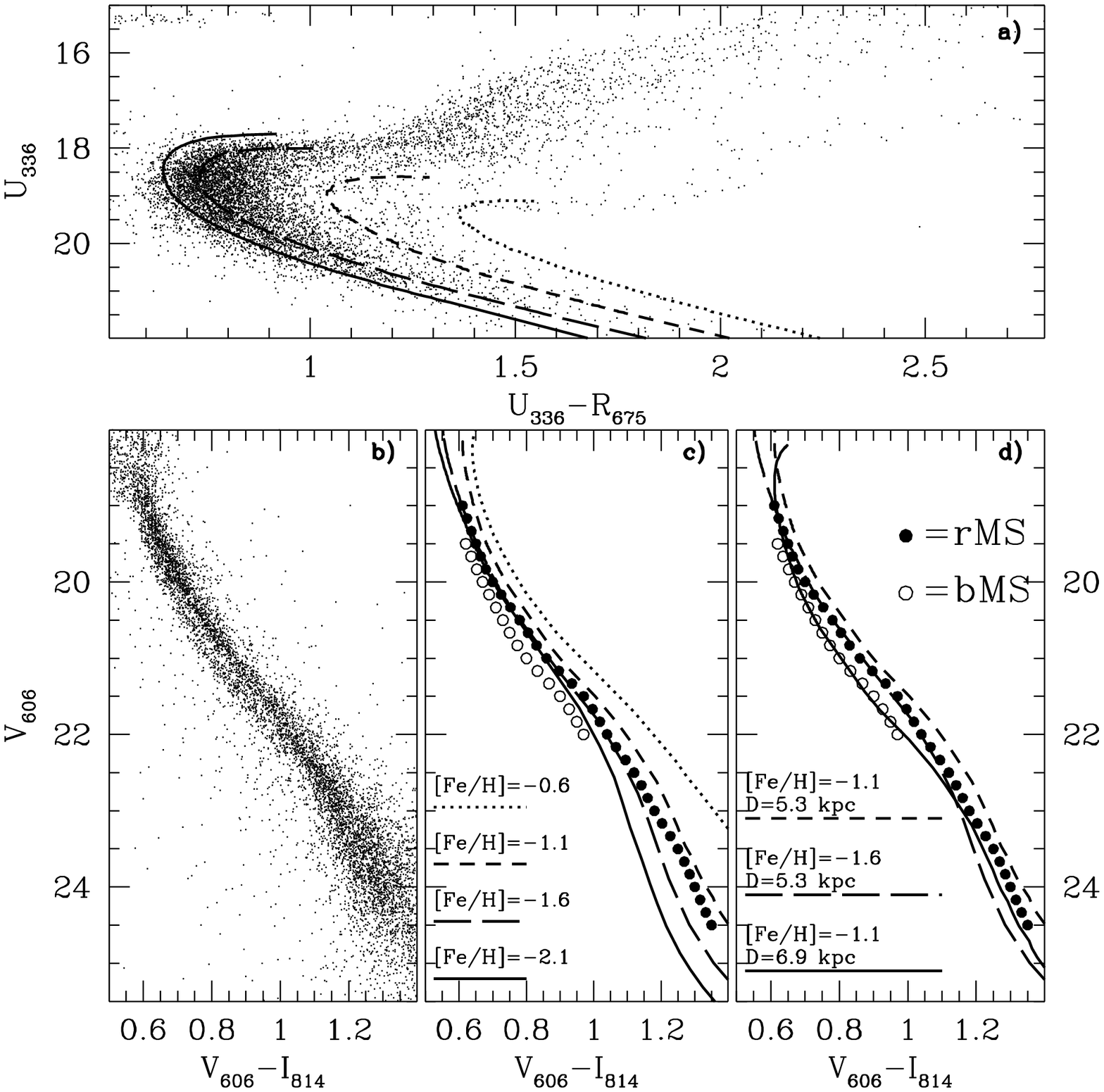}
\caption{Comparison of  theoretical models with  WFPC2 observations of
the TO region (panel a), and the DMS adopting the same distance (as in
panel a) for  all the  isochrones (panel c),  and  shifting the [Fe/H]
$\!=\!-1.1$ isochrone  by 1.6  kpc  (panel  d).  For  clarity we  have
plotted only the fiducial points of the rMS and the bMS. }
\end{figure}


\clearpage

\begin{thebibliography}{}
%
\bibitem[]{524} Allard, F., Hauschildt, P.\ H., Alexander, D.\ R., \& Starrfield, S. 1997, \araa, 35, 137
\bibitem[]{525} Anderson, J. 1997, PhD Thesis, UC, Berkeley 
\bibitem[]{526} Anderson, J., \& King, I.\ R. 2000, \pasp, 112, 1360
\bibitem[]{527} Anderson, J. 2002, in ASP Conf.\ Ser.\ 265,  
                 {\it ``$\omega$ Centauri a Unique Window into Astrophysics''}, 
                 ed.\ F.\ van Leeuwen, J.\ Hughes,  \& G.\ Piotto  (San Francisco: ASP), 87
\bibitem[]{530} Anderson, J. 2003, in ASP Conf. Ser. 296, 
                 {\it ``New Horizons in Globular Clusters Astronomy''},
                 ed.\ G.\ Piotto, G.\ Meylan, S.\ G. Djorgovski, \& M.\ Riello (San Francisco: ASP), 125
\bibitem[]{533} Cassisi, S., Castellani, V., Ciarcelluti, P., Piotto, G., \& 
                 Zoccali, M. 2000, \mnras, 315, 679
\bibitem[]{535} Cassisi, S., Castellani, V., Degl'Innocenti, S., Salaris, M., \& 
                 Weiss, A.  1999, \aaps, 134, 103
\bibitem[]{550}	D'Antona, F., Caloi, V., Montalb\'an, J., Ventura, P., \& Gratton, R. 2002, \aap, 395, 69
\bibitem[]{539} Ferraro, F.\ R., Bellazzini, M., \& Pancino E. 2002, \apj, 573, L95
\bibitem[]{540} Ferraro, F.\ R., Sollima, A., Pancino E., Bellazzini, M., 
                 Straniero, O., Origlia, L., \& Cool, A.\ M. 2004, preprint (astro-ph/0401540)
\bibitem[]{542} Hilker, M., \& Richtler, T. 2000, \aap, 362, 895
\bibitem[]{543} Holtzman, J.\ A., Burrows, C.\ J., Casertano, S., Hester, J.\ J.,
                 Trauger, J.\ T., Watson, A.\ M., \& Worthey, G. 1995, \pasp, 107, 1065
\bibitem[]{545} Jurcsik, J. 1998, \apj, 506, L113
\bibitem[]{546} Lee, Y.\ W., Joo, J.\ M., Sohn, Y.\ J., Rey, S.\ C., 
                 Lee, H.\ C., \& Walker, A.\ R. 1999, \nat, 402, 55
\bibitem[]{548} Leon, S., Meylan, G., \& Combes, F. 2000, \aap, 359, 907
\bibitem[]{549} Norris, J.\ E., \& Da Costa, G.\ S. 1995, \apj, 447, 680
\bibitem[]{564} Norris, J.\ E., Freeman, K.\ C., \& Mighell, K.\ J. 1996, \apj, 462, 241
\bibitem[]{551} Origlia, L., Ferraro, F.\ R., Bellazzini, M., \& Pancino, E. 2003, \apj, 591, 916
\bibitem[]{552} Pancino, E., Ferraro, F.\ R., Bellazzini, M., Piotto, G., \& 
                 Zoccali, M. 2000, \apj, 534, L83
\bibitem[]{554} Piotto, G., \& Zoccali, M. 1999, \aap, 345, 485
\bibitem[]{555} Platais, I., Wyse, R.\ F.\ G., Hebb, L., 
                 Lee, Y.\ W., \& Rey, S.\ C. 2003, \apj, 591, L130
\bibitem[]{572} Salaris, M., Riello, M., Cassisi, S., \& Piotto, G.  2004, \aap, subm.
\bibitem[]{} Smith, V.\ V., Suntzeff, N.\ B., Cunha, K., Gallino, R., Busso, M., Lambert, D.\ L., 
                 \& Straniero, O., \aj, 119, 1239.
%
\end{thebibliography}
\end{document}